\documentclass[aps,prl,twocolumn,showpacs,preprintnumbers,superscriptaddress]{revtex4}
\usepackage{graphicx}
\usepackage{epsf} 
\usepackage{dcolumn}
\usepackage{bm}
\usepackage{amssymb}
\usepackage{amsmath}

\begin{document}

\title{Dynamic Scaling of Bred Vectors in Chaotic Extended Systems}

\author{Cristina Primo}

\affiliation{Instituto de F\'{\i}sica de Cantabria (CSIC--UC),
E-39005 Santander, Spain} \affiliation{Instituto Nacional de
Meteorolog{\'\i}a CMT/CAS, Santander, Spain}

\affiliation{Departamento de Matem\'atica Aplicada, Universidad de
Cantabria, Avda. Los Castros, E-39005 Santander, Spain}

\author{Miguel A. Rodr{\'\i}guez} 

\author{Juan M. L{\'o}pez} 
\affiliation{Instituto de F\'{\i}sica de Cantabria (CSIC--UC),
E-39005 Santander, Spain}

\author{Ivan Szendro} 

\affiliation{Instituto de F\'{\i}sica de Cantabria (CSIC--UC),
E-39005 Santander, Spain}

\affiliation{Departamento de F{\'\i}sica Moderna, Universidad de
Cantabria, Avda. Los Castros, E-39005 Santander, Spain}
\date{\today}

\begin{abstract}
We argue that the spatiotemporal dynamics of bred vectors in
chaotic extended systems are related to a kinetic roughening
process in the Kardar-Parisi-Zhang universality class. This
implies that there exists a characteristic length scale
corresponding to the typical extend over which the finite-size
perturbation is actually correlated in space. This can be used as
a quantitative parameter to characterize the degree of projection
of the bred vectors into the dynamical attractor.
\end{abstract}


\maketitle

A standard tool for studying chaotic behavior in dynamical systems
is the computation of the characteristic Lyapunov exponents,
which, roughly speaking, measure the typical exponential growth
rate of an infinitesimal disturbance \cite{schuster,ott}. The
characteristic Lyapunov exponents in extended systems are defined
in a similar way as their low-dimensional counterpart and can be
calculated from the linearization of the equations of motion
\cite{boffetta,bohr}. The main point is that the growth of an
infinitesimal perturbation is described by the linear equations
for the tangent space, the so-called Lyapunov vectors. However,
for many practical purposes, Lyapunov vectors may be irrelevant as
indicators of, for instance, the predictability time. Indeed, in
realistic situations, the error in the initial condition is
finite. The important fact is that the evolution of finite errors
is not confined to the tangent space, as defined by the growth of
linearized perturbations, but is controlled by the complete
non-linear dynamics. In order to deal with realistic
perturbations, the concept of finite size Lyapunov exponents has
been found \cite{aurell} to be useful to analyze predictability in
high-dimensional systems \cite{boffetta}.

A good example with important practical application occurs in
weather forecasting research, where one deals with the whole
Earth's atmosphere-- an extremely high dimensional system in which
initial conditions can be determined only with limited accuracy.
In this context, Patil {\it et. al.} have recently introduced the
concept of {\em bred vector} (BV) \cite{patil,kalnay}, as the
spatio-temporal evolution of a statistical ensemble of
non-infinitesimal perturbations. From the analysis of real data
provided by the US National Weather Service, Patil {\it et. al.}
have proposed and measured an effective local finite-time
dimension, by means of the so-called local BV dimension statistic
\cite{patil,kalnay}. Regions of low BV dimension are identified as
more predictable than locations where dimension is high. The
method has recently been applied to 2D coupled map lattices in
which nonlinear time series analysis has been used to make local
predictions of trajectories at low dimension sites
\cite{francisco}. Breeding techniques constitute an important tool
in modern weather forecasting research \cite{kalnay} and one can
envisage that BVs can be used in other spatially extended systems
with chaotic dynamics where local short time forecasts can also be
feasible \cite{francisco}. However, many questions concerning
statistical and dynamical properties of BVs are still unanswered.
The aim of this Letter is to shed some light into these questions.

In this Letter we study the spatiotemporal dynamics of homogeneous
finite-size errors and focuss on the propagation dynamics of BVs
in chaotic extended systems. We argue that, after a suitable
transformation of variables, the BV dynamics can be interpreted as
a kinetic roughening process in the Kardar-Parisi-Zhang
universality class \cite{kpz}. We also find that the breeding
procedure introduces a characteristic length scale corresponding
to the typical extend over which the finite-size perturbation is
actually correlated in space. This can be used as a quantitative
parameter to characterize the degree of projection of the BVs into
the dynamical attractor, which is of major interest in
probabilistic forecasting techniques based on ensembles of BVs.

We exemplify our results by means of numerical simulations of
coupled map lattices in one dimension, which are simple model
systems exhibiting space-time chaos and convenient as far as the
computing time is concerned. Then, we shall be considering a
coupled map array consisting of $L$ chaotic oscillators given by
\begin{eqnarray}
u(x,t+1)= \nu \,f(u(x+1,t))\,+
\nonumber\\
\nu\, f(u(x-1,t))\,+\,(1-2\,\nu)\,f(u(x,t)) \label{cml}
\end{eqnarray}
where $x = 1, 2, \ldots, L$, $f(u)$ is a chaotic map, $\nu$ is the
coupling constant, and periodic boundary conditions are imposed.
We have fixed the coupling to $\nu = 1/3$ in all the simulations
presented in this Letter. We have carried out simulations for two
different choices of the map, the chaotic logistic map $f(u) =
4u(1-u)$, $0 \le u \le 1$ and the tent map $f(u) = 1 - 2\vert
u-1/2\vert$, $0 \le u \le 1$. For the sake of brevity, all the
results we present below correspond to coupled logistic maps, but
similar results were obtained for the tent map.

\paragraph{Scaling of finite perturbations.--}
Let us first consider the evolution of finite size homogeneously
perturbed trajectories in our model system (\ref{cml}). Given an
initial condition $u^0(x,0)$, the solution $u^0(x,t)$ is
univocally determined by computing Eq.(\ref{cml}) for a number $t$
of time steps. This will be our reference trajectory and we shall
be studying the evolution of finite perturbations around that
reference solution. Since we are interested here in the
propagation of real (non-infinitesimal) errors, we should avoid
linearization of Eq.(\ref{cml}). Instead, we compute the
trajectories generated by iterating (\ref{cml}) for an ensemble of
initial conditions $u(x,0) = u^0(x,0)+\delta u(x,0)$, where
$\delta u(x,0)$ is uniformly distributed in
$(-\epsilon_0,\epsilon_0)$. For each iteration of the lattice
(\ref{cml}) the difference $\delta u(x,t)=u(x,t)-u^0(x,t )$
between the reference trajectory and every one of the disturbed
solutions is calculated and represents the evolution of
homogeneously distributed finite errors in the system. Although
the disturbances are initially homogeneous and uncorrelated in
space, as time goes by, they propagate, get correlated, and grow
exponentially in size. The statistical fluctuations of
disturbances can be characterized by studying now the ensemble of
finite perturbations $\{\delta u_{n}(x,t)\}_{n=1}^{N}$, which
correspond to $N$ independent realizations of the initial
perturbation. In our simple model we can use ensembles of hundred
of samples to obtain a good statistics.
\begin{figure}
\centerline{\epsfxsize=7.5cm \epsfbox{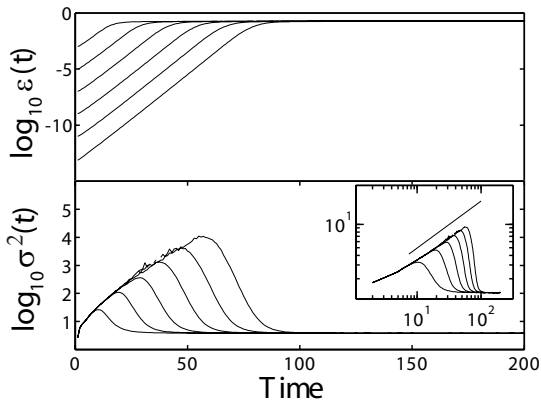}}
\caption{Numerical results for the propagation of finite-size
errors in the coupled logistic maps. Upper panel shows the
amplitude factor $\epsilon(t)$ {\it vs.} time for perturbations
starting with initial amplitudes of $\epsilon_0 = 10^{-3},
10^{-5}, 10^{-7}, 10^{-9}, 10^{-11}$ and $10^{-13}$ (from top to
bottom) in 1D lattices of $L=1024$ sites and results were averaged
over 600 different initial conditions. Lower panel shows the
variance $\sigma^2(t)$ for the same initial perturbations as
before and $\epsilon_0$ decreasing from top to bottom. The same
data are plotted in log-log scale in the inset to show the
power-law behavior $\sigma^2(t) \sim t^{2\beta}$. The straight
line has a slope $0.60$ and is plotted to guide the eye.}
\end{figure}

Firstly, we introduce what we call the {\em amplitude factor}
$\epsilon(t)$ as the spatial geometrical mean value of the
perturbation:
\begin{equation}
\epsilon(t) \equiv \prod_{x=1}^L \vert \delta u(x,t)\vert^{1/L},
\label{amplitude}
\end{equation}
which, as we shall see below, contains the information about the
dominant exponential growth rate. In Fig.\ 1 we plot $\log
\epsilon(t)$ {\it vs.} time for different values of the initial
perturbation amplitude $\epsilon_0$. We can see that for times $t
< t_{\rm lin}(\epsilon_0)$ the average amplitude factor grows
exponentially in time $\epsilon (t) \approx \epsilon_{0}
\exp(\lambda t)$. We demonstrate below that $\lambda$ indeed
corresponds to the maximal Lyapunov exponent. For longer times, $t
> t_{\rm lin}(\epsilon_0)$ the amplitude factor saturates to a
constant value. Both, the saturation constant and the maximal
Lyapunov exponent, are independent of the initial perturbation
size $\epsilon_0$. However, the saturation times $t_{\rm
lin}(\epsilon_0)$ increase as the size of the initial perturbation
$\epsilon_0$ becomes smaller. This indicates that $t_{\rm lin}$
corresponds to the crossover time at which the dynamics of a
finite size perturbation depart from the linear approximation
({\it i.e.} the Lyapunov vectors). This crossover occurs because
the tangent Lyapunov vectors describe only the behavior of
strictly infinitesimal perturbations ($\epsilon_0 \to 0$).
Therefore, for times $t > t_{\rm lin}(\epsilon_0)$ nonlinear
corrections, due to finiteness of the initial perturbation, come
into play and force the errors out of the tangent space. From then
on, the linear approximation cannot describe the evolution of
errors. One then expects that $t_{\rm lin}(\epsilon_0) \to \infty$
as $\epsilon_0 \to 0$.

Besides exponential growth, correlations are dynamically generated
during the evolution of perturbations. Correlations contain
information about the sub-leading Lyapunov exponents and thus also
contribute to the perturbation size growth. The important role of
correlations can be better realized after subtraction of the
dominant exponential growth component given by $\epsilon(t)$. We
find that a very useful indicator is given by the {\em reduced}
perturbations $\delta r(x,t)$ that we define as
\begin{equation}
\delta r(x,t)=\frac{\delta u(x,t)}{\epsilon(t)}, \label{reduced}
\end{equation}
where the dominant exponential growth is globally removed and one
is left with the effect of correlations. Statistical fluctuations
of the reduced perturbations are measured by the variance
$\sigma^2(t) = \langle \overline{\delta r(x,t)^2} \rangle$, where
$\langle \cdots \rangle$ stands for average over realizations of
the initial perturbation and the over bar is a spatial average. As
we shall see below, $\sigma(t)$ gives information about the growth
of perturbations due solely to correlations.

In Fig.\ 1 we show our numerical results for the variance of the
reduced perturbations, Eq.(\ref{reduced}). The variance
$\sigma^2(t)$ grows as an exponential power law $\sigma^2(t)\sim
\exp(t^{2\beta})$ for times $t < t_{\rm lin}$, {\it i.e.} before
saturation by finiteness of the initial disturbance. A
least-squared fit of $\log(\log(\sigma))$ {\it vs.} $\log(t)$
gives the exponent $\beta = 0.30 \pm 0.05$ (see the inset of Fig.\
1). As before, this rapid growth occurs for times such that the
dynamics of disturbances are well described by the Lyapunov
vectors.

The magnitude and extent of spatial correlations can be measured
by use of the site--site correlation function $\rho(x,t) = \langle
\overline{\delta u(x_0,t) \delta u(x+x_0,t)}\rangle/\langle
\overline{\delta u(x_0,t)^2}\rangle$. The magnitude of spatial
correlations increases in time $\rho(x,t_1) < \rho(x,t_2)$ if $t_1
< t_2$ for times $t_1, t_2 < t_{lin}$. However, correlations
become progressively smaller for times $t > t_{lin}$. This
indicates that the effect of having initially finite perturbations
is to introduce a characteristic time $t_{lin}(\epsilon_0)$
marking the typical time it takes for the system to depart from
tangent space (with the building-up of correlations) to truly
non-linear evolution (uncorrelated errors). For times larger than
$t_{lin}$ perturbations quickly get spatially uncorrelated, as can
be seen from the decay of the reduced variance in Fig.\ 1. Further
analysis \cite{cris-unpub} shows that errors become
undistinguishable from actual white noise.

Our numerical results can be explained analytically by making use
of a transformation first proposed by Pikovsky and Politi for the
actual Lyapunov vectors \cite{pik1,pik2}. We find that the
dynamics of finite perturbations can also be seen as a kinetic
roughening processes of the surface defined by $h(x,t) = \log
\vert \delta u(x,t) \vert$. This is a very useful transformation
that allows us to make use of existing results in the field of
nonequilibrium surface growth. Indeed, Politi and Pikovsky
\cite{pik1,pik2} have shown that errors in many extended systems
lead to surface growth process in the universality class of KPZ
\cite{kpz}. Let us first consider the amplitude factor defined in
Eq.\ (\ref{amplitude}) and show that it is related to the average
surface velocity. We can write $\vert \delta u(x,t) \vert = \exp[
h(x,t)]$ and thus $\epsilon(t) = \exp[(1/L)\sum_{x=1}^{x=L}
h(x,t)] = \exp[\overline{h}(t)]$. We then obtain that the
amplitude factor must grow as $\epsilon(t) = \epsilon_0
\exp(\lambda t)$, where $\lambda$ is the largest Lyapunov exponent
and corresponds to the surface velocity, in agreement with our
numerical results shown in Fig.\ 1.

Also the time behavior of the variance of the reduced
perturbations shown in Fig.\ 1 can be analytically related to
surface scaling properties. We find that $\sigma(t)^2 = \sqrt{2
\pi} \exp[2 W(L,t)^2]$, where $W(L,t)$ is the surface width and
depends on the system size $L$. This result is easily obtained by
assuming a Gaussian distribution of surface heights. KPZ behavior
then implies that the width scales as $W(L,t) \sim t^\beta$ for
times $t < t_s(L)$ and saturates to a size dependent value,
$W(L,t) \sim L^\alpha$ for $t > t_s(L)$, where $\beta$ and
$\alpha$ are the growth and roughness exponent respectively that
are known to have the values $\beta=1/3$ and $\alpha = 1/2$ for
the KPZ universality class in one dimension \cite{kpz}. This
explains the $\exp(t^{2/3})$ scaling observed in the inset of
Fig.\ 1. It is worth mentioning that all the numerical results
presented here are obtained for system sizes such that $t_{\rm
lin} \ll t_s$, so that the existence of $t_{\rm lin}$ could be
clearly seen.

\paragraph{Dynamic scaling of bred vectors.--} An important
problem where finite initial errors are a concern is in weather
forecasting. In this context, initial configurations that belong
or are very close to the dynamical attractor are much sought
after. These prepared configurations are then used as initial
conditions in the atmospheric models in order to have a
probabilistic forecast for a certain time window. Predictions are
expected to be better when the initial conditions are closer to
the dynamical attractor of the system. To achieve the goal of a
maximal projection, the computer runs of the atmospheric models
are started from random perturbations and BV techniques are used
\cite{kalnay} to allow the perturbed trajectories to get closer to
the chaotic attractor. BVs are defined in analogy to the operation
of data reassimilation in numerical models of atmospheric
evolution, in which the output of the numerical model is corrected
by the observed experimental data after short periods of time
$\{\tau _{1} ,\tau _{2} , \tau _{3}.. \}$. BVs are defined by
multiplying each member of the ensemble by a reduction factor
$\{k_{1} , k_{2}, k_{3}...\}$ at those times so that the amplitude
factor of the perturbed solutions $\epsilon_0$ is kept constant.
Several criteria could be chosen to construct BVs. We define here
what we call {\em continuous bred vectors} $\delta B (x,t)$ as the
perturbation obtained by dividing by the amplitude factor at every
time step. Since the dominant exponential growth is filtered out,
BVs evolve with a constant amplitude factor by definition, but
exhibit the actual spatial correlations (see \cite{kalnay} for
further details).

We have studied the dynamics of BVs in the coupled-map lattice
model given by Eq.\ (\ref{cml}). As in the case of finite
perturbations we have considered a reference trajectory $u^0(x,t)$
and its corresponding ensemble of randomly perturbed trajectories
$u(x,0) = u^0(x,0)+ \delta u(x,0)$. But now breeding is applied so
that we define the re-scaled error (the BV) $\delta
\mathcal{B}(x,t) = \delta u(x,t)/\epsilon(t)$ at every time step
and the perturbed trajectories are then evolved by introducing
$u^0(x,t) + \epsilon_0\, \delta \mathcal{B}(x,t)$ in Eq.\
(\ref{cml}) to obtain $u(x,t+1)$. We have carried out simulations
of the model for different values of the finite amplitude
$\epsilon_0$, which is the only external parameter. The
implemented procedure is intended to mimic the one used in
probabilistic weather forecasting \cite{kalnay}.

The reduced perturbations $\delta r(x,t) = \epsilon_0\, \delta
\mathcal{B}(x,t)$ and, in particular its variance, give a direct
measure of the spatial fluctuations of the BVs. As we have shown
in our previous analysis of finite perturbations, which also
applies here, the variance $\sigma^2(t)$ grows with time as
$\sigma^2\sim \exp(t^{2\beta})$ until it saturates due to
nonlinear terms becoming important in the equation of evolution of
finite errors. The main point now is that, in contrast with the
case of bare finite-size perturbations, the breeding procedure
leads to a non-trivial stationary value of the statistical
fluctuations of BV. More importantly, this stationary state is
characterized by having space correlations, as can be seen from
the finite stationary values of the variance $\sigma^2$ in Fig.\
2. The time scale $\tau(\epsilon_0,L)$ characterizes the crossover
to the stationary regime, where fluctuations are no longer time
dependent. After that time BVs are correlated over regions of
characteristic size $l(\epsilon_0,L) \sim \tau^{1/z}$, where $z =
\alpha/\beta$ is the dynamic exponent. We find that numerical data
of BV fluctuations can be cast in a {\em dynamic scaling ansatz}
given by
\begin {equation}
\log[\sigma^2(t,\epsilon_0)]= t^{2\beta}\,
\mathcal{G}(t/\tau(\epsilon_0)) \label{dyn-scal},
\end {equation}
where the scaling function $\mathcal{G}(v) \sim v^{-2\beta}$ if $v
\gg 1$, and $\mathcal{G}(v) \sim {\rm const}$ if $v \ll 1$. The
characteristic time is found to scale as $\tau(\epsilon_0) \sim
\vert\log(\epsilon_0) \vert^{1/\beta}$, as shown by collapse of
the data in the inset of Fig.\ 2.
\begin{figure}
\centerline{\epsfxsize=7.5cm \epsfbox{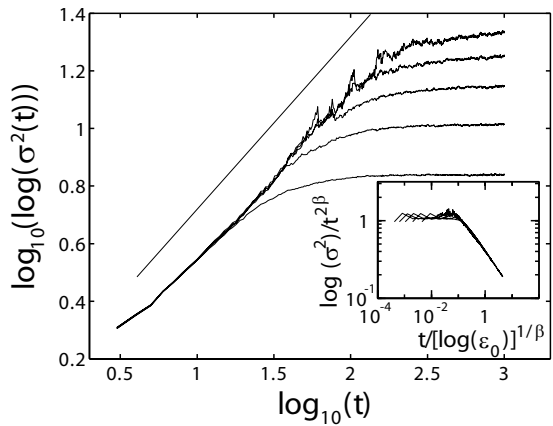}}
\caption{Dynamic scaling of BVs for the coupled logistic maps.
Main panel shows the variance $\sigma^2(t)$ {\it vs.} time for
perturbations kept at finite amplitudes $\epsilon_0 = 10^{-3},
10^{-4}, 10^{-5}, 10^{-6}$ and $10^{-7}$ (from bottom to top) by
means of the breeding procedure as described in the text. The
straight line has a slope 0.60 and is a guide to the eye. The
inset shows a data collapse in log-log scale according to Eq.\
(\ref{dyn-scal}) with exponent $\beta = 0.28 \pm 0.05$. Results
were obtained in systems of linear size $L=1024$ and averages over
500 different initial conditions were taken.}
\end{figure}

We can now make use of the surface growth mapping, $h(x,t) = \log
\vert \delta \mathcal{B}(x,t) \vert$, to obtain some analytical
understanding on the dynamic scaling behavior of BVs. Among other
quantities, one can obtain the scaling law for the saturation time
$\tau(\epsilon_0,L)$ as follows. Let $M$ be the upper bound for
the linear approximation to be valid to describe the evolution of
perturbations in the system. To be precise, if the error is larger
than $M$,  $\vert \delta u \vert  > M$, then the time evolution of
$\delta u(x,t)$ is not longer in the tangent space
(infinitesimal), but finite and to be described by higher order
nonlinearities. Any finite-size perturbation in the linear regime
is then bounded by $\epsilon_0 \,\delta \mathcal{B} < M$. This
implies an upper bound in the value that the surface height can
take $h_{max} \sim \vert\log(M/\epsilon_0)\vert$ and therefore a
saturation time $\tau$ corresponding with the typical time for
which the surface width $W(\tau) \sim h_{max}$. This leads to a
characteristic time scale $\tau(\epsilon_0) \sim
\vert\log(M/\epsilon_0)\vert^{1/\beta}$. This time scale together
with the usual saturation time $\tau_c \sim L^z$ associated with
the kinetic roughening of the surface $h$ are the two
characteristic time scales in the system. Saturation of the
surface fluctuations is then controlled by the shortest of the
two, $\tau(\epsilon_0,L) \sim \min
\{\vert\log(M/\epsilon_0)\vert^{1/\beta}, L^z \}$. This
theoretical argument agrees very well with our numerical data as
can be seen in the inset of Fig.\ 2, where an excellent data
collapse is obtained by using our theoretical expression for the
saturation time $\tau \sim
\vert\log(M/\epsilon_0)\vert^{1/\beta}$. For clarity of
presentation all the values of $\epsilon_0$ shown in Fig.\ 2
satisfy $\vert\log(M/\epsilon_0)\vert^{1/\beta} \ll L^z $ so that
saturation time is governed by
$\vert\log(M/\epsilon_0)\vert^{1/\beta}$. On the contrary if
smaller systems are used one can see that $\tau \sim L^z$
\cite{cris-unpub}.

Due to the scale--invariant dynamics of the surface $h(x,t) = \log
\vert \delta \mathcal{B}(x,t) \vert$ there exists a length scale
$l(\epsilon_0,L) \sim \tau(\epsilon_0,L)^{1/z}$ that corresponds
to the typical extend of spatial correlations in the stationary
regime. This length scale is a quantity of great importance since
it can be used as an indicator of the degree of projection of the
BVs into the dynamical attractor. If $l < L$ correlations are
short ranged and the perturbed trajectory is only partially
projected over the attractor. On the contrary, for $l > L$ BVs are
spatially correlated over whole system and projection is maximal.
In this case, BVs dynamics are confined to the tangent space.

Finally, we can draw some interesting conclusions for the
generation and application of BVs as ensembles of initial
conditions in probabilistic forecasting. For a given system size
$L$ there is a threshold $\epsilon_{th} \sim \exp(L^{\alpha})$ of
the fixed amplitude of BVs such that disturbances with amplitudes
$\epsilon_0 < \epsilon_{th}$ propagate as infinitesimal
disturbances, which dynamics are fully described by the tangent
space equations (Lyapunov vectors). This in turn means that BVs
are then maximally projected and correlated over whole system. In
contrast, BVs with $\epsilon_0
> \epsilon_{th}$ are non infinitesimal disturbances that cannot be
described by tangent space equations, but by the full nonlinear
dynamics. In this case projection into the attractor is only
partial and correlations have a characteristic size
$l(\epsilon_0,L)$. Our results show that when using breeding
techniques in probabilistic forecasting for the preparation of
ensembles of initial conditions, the election of the normalization
condition of the BVs is of vital importance since it introduces a
previously unforeseen time/length scale related to degree of
projection into the dynamical attractor.

\begin{acknowledgements}
Financial support from the Ministerio de Ciencia y Tecnolog{\'\i}a
(Spain) under project BFM2000-0628-C03-02 is acknowledged.
\end{acknowledgements}

\end{document}